\documentclass[floatfix,aps,prl,amsmath,nofootinbib,
preprintnumbers,twocolumn]{revtex4}
\usepackage{graphicx}

\begin{document}

 \preprint{ITPCAS-PUB-05-03-A}

 \title{Super-horizon Perturbations and CMBA}
 \author{Ding-fang Zeng and Yi-hong Gao}\email{dfzeng@itp.ac.cn, gaoyh@itp.ac.cn}
 \affiliation{Institute of Theoretical Physics of Chinese Academy
 of Science, Beijing, China}

 \date{\today}

 \begin{abstract}
 We provided a gedanken experiment and
 argued that since observers inside a given Hubble volume could not detect the
 super horizon perturbation modes as
 real perturbations, these modes could only affect the average
 value of the cosmic microwave background (CMB), but not its
 anisotropy
 properties (CMBA) in that Hubble volume.
 \end{abstract}


 \maketitle

 \newcommand{\beq}[1]{\begin{eqnarray}\label{#1}}
 \newcommand{\eeq}{\end{eqnarray}}


 In \cite{KMNR},
 it is proposed that even in a totally matter filled universe,
 the accelerating expansion phase can be observed because
 observers in a given Hubble volume cannot detect the
 super-Hubble
 perturbations as real perturbations, instead they could only detect them
 as time-dependent background in that Hubble volume.
 By this reasoning, when CMB is
 concerned,
 the super-horizon perturbation modes could only affect the
 average value but not the anisotropic properties of it. There is a
 vivid gedanken experiment helping us to understand this fact.
 Imagine that you were put on a closed ship in a lake which was
 waving, after a period of time, you will find that you were put
 in a un-flat environment. But if the wave which was hitting your ship
 had period longer than the time you had been in the
 ship, you would not find the environment was un-flat!

 From the aspects of statistic physics, CMBA \cite{Dod03} is just the manifestation of photon
 distribution's deviation from exact Bose-Einstein formulaes,
 \beq{}
 f(t,\vec{x},p,\hat{p})=\frac{1}{\textrm{exp}[\frac{p}{(1+\Theta(t,\vec{x},\hat{p}))T}]-1}
 \label{NonBEdistribution}
 \eeq
 Using Fourier expansion,
 \beq{}
 \Theta(t,\vec{x},\hat{p})&&\hspace{-3mm}=\int
 d\vec{k}\textrm{e}^{i\vec{k}\cdot\vec{x}}\Theta(t,\vec{k},\hat{p})
 \label{FourierExpandingTheta}
 \eeq
 Mathematically, the range of integration in
 eq(\ref{FourierExpandingTheta}) is $(-\infty,\infty)$.
 However, for perturbations
 with wave length greater than the current Hubble radius,
 observers inside the Hubble volume could not detect them as real
 perturbations, they can only look them as background. These perturbation
 modes may make the average temperature of CMB in one
 Hubble volume different from that in another, but they do
 not affect its anisotropy properties in a
 given Hubble volume. Appropriately, when
 calculating the angular power spectrum of CMBA \cite{Dod03, Seljak},
 contributions from these perturbation modes should
 also be excluded, i.e.,
 \beq{}
 C_l=4\pi\int_{H_0}^{\infty} dkk^2P_{\Psi}(k)\Theta_l^2(t,k)
 \label{Correlation2Cal}
 \eeq
 where $P_{\Psi}(k)$ is the primordial power spectrum of
 perturbations produced during inflation, $\Theta(t,k)$ describe
 the time evolution of this power spectrum.
 \begin{figure}
 \includegraphics[scale=0.4]{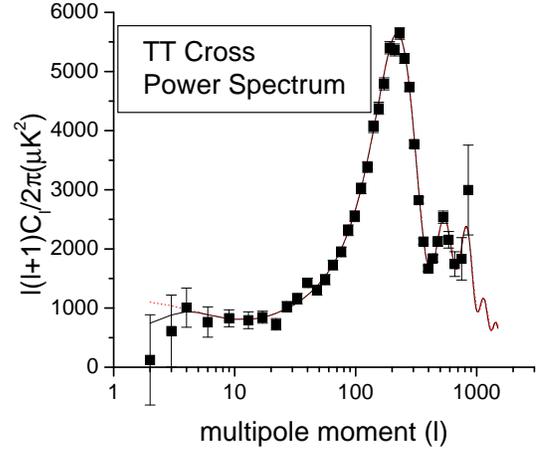}
 \caption{
 The angular power spectrum of CMBA. The scattered
 point is the observational results of \cite{WMAP1},
 the solid curve is the result of
 eq(\ref{Correlation2Cal}), the dotted curve is
 calculated without $k$ cut-off in the integration, both curves
 are calculated by CMBFast4.5.1 \cite{Seljak} with the best
 fitting cosmic parameters of \cite{WMAP1} as input. The input
 amplitude of the primordial power spectrum which gives the
 solid line is amplified to give the same maximum multi-pole
 moment as that of the dotted line.
 }\label{cltt}
 \end{figure}
 In FIG.1, we compared the angular power spectrum of CMBA calculated from
 eq(\ref{Correlation2Cal}) and the usual zero cut-off case. From
 the figure we see that, imposing a cut off although may not
 definitely address the little $l$ problem\cite{CPLL}, it ameliorate the
 problem at least. In \cite{DG1} we also expressed relevant
 ideals.

 Superficially, according to our gedanken experiment, to detect
 the un-uniformity of CMB at the scale of $H_0^{-1}$, we have to start our
 experiment as soon as the universe is born. But, in practice,
 what we would detect is the correlation
 $<\Theta(t,\vec{x},\hat{p}_1)\Theta(t,\vec{x},\hat{p}_2)>$ instead $\Theta(t,\vec{x},\hat{p})$
 itself. This makes it possible to detect the un-uniformity signal
 even in a few seconds! But the fact that photons
 carrying signals which will be used to uncover the super-horizon un-uniformity
 have not arrived our detector
 cannot be changed by techniques.


\end{document}